\magnification=1200
\def\build#1_#2^#3{\mathrel{\mathop{\kern 0pt#1}\limits_{#2}^{#3}}}
\catcode `@=11

\def\n{\noindent}
\def\m{\medskip}

\hsize 17truecm
\vsize 24truecm

\baselineskip 18pt
\nopagenumbers

\font\tenll=lasy10
\newfam\llfam
\textfont\llfam=\tenll

\parskip=6pt plus 2pt minus 2pt

\baselineskip 12pt

\vskip 1.5truecm
  \centerline{\bf  QUANTUM GRAVITY: THE AXIOMATIC APPROACH,}
\bigskip 
\centerline{\bf A POSSIBLE INTERPRETATION.}
\bigskip 
  
\vskip 1truecm
\centerline{\bf Martine PERRIN, Guy BURDET} 

\medskip
\centerline{Centre de Physique Th\'eorique, CNRS Luminy}
\smallskip
\centerline{Case 907, F-13288 Marseille cedex 9, France}
\smallskip
\centerline{email: perrin@cptsu2.univ-mrs.fr}

\vskip 2truecm
\noindent { \bf Abstract}   
\vskip .5truecm
Twenty years ago, by extending the Wightman axiom framework, it has been found possible
 to quantize only a conformal factor of the gravitational field. Gravitons being excluded
from this quantum scalar field theory, numerous attempts were done to give a valuable
description of what could be quantum gravity. In this talk we present a familly of Lorentz
manifolds which can be foliated by isotropic hypersurfaces and pose severe restrictions on
the form of the energy-momentum tensor in Einstein's equations. They can be associated to
gravitational waves "without gravitons" in a vacuum described by two cosmological functions,
but not to a massless particle flow. From this cross-checking with the previous remark, a
"very" primordial quantum cosmological scenario is proposed. \bigskip \bigskip \bigskip \bigskip 
$\it ... \ it \ is \ important \ that \ everyone \ does  \ \hbox {\bf not } \ follow \ the \ same \
approach,  \ work \ with \ the $

$ same \ set \ of \ prejudices. \ In \  a \ field \ like \ quantum \ gravity \
where  \ the \ experimental $

$ data \ is \ so \ scarse, \ diversity \ of \ ideas \ is \ all \ the \
more  \ important. \ The \ advice \ that $

$ Richard \ Feynman \ gave \ to \ younger \ researchers \ at \
CERN \ on \ his \ way \ back \ from $

$ Stockholm \ in \ 1965 \ seems \ especially \ apt \ in \ the \ context \ of \ quantum \ gravity: $

{\leftskip=2cm \noindent  It's very important that we do not all
follow the same fashion. ...It's necessary to increase the amount of variety ... and
the only way to do it is to implore you few guys to take a risk with your lives that
you will be never heard of again, and go off in the wild blue yonder and see if you can
figure it out. \par } 

{ \leftskip=8,1cm $A.Ashtekar, \ Les Houches, \ 1992.$\par }
\vfill \eject

One of the long-standing challenges of contemporary theoretical physics is the unification of
Einstein's general relativity theory with quantum theory, an unsolved problem for over sixty
years. If a considerable amount of progress has been made in our understanding of quantum
processes occuring in a strong gravitational field, a quantum theory of the gravitational 
field itself still does not exist.

After the astonishing ability of the quantum electrodynamics to survive, the fifties ended
over a great hope, the axiomatic approach of the general theory of quantized fields known as
the Wightman's axioms that one briefly recalls to be self-consistent.

\n{\bf I. The axiomatic field theory} (for a neutral scalar field) [1].

\n$ Zeroth\ axiom$ : The $space \ of \ states$ is the set of unit rays in a complex Hilbert
space ${\cal H}$. \m \n $ First\ axiom$ : The $scalar \ quantum \ field$ is an operator
valued tempered distribution on the Minkowski spacetime $\bf R^{3,1}$ i.e. ${\cal L} \in
{\cal S'}({\bf R^{3,1}} ,  op({\cal H}))$. The set ${\cal S}$ of test functions
($C^{\infty}$-functions $f$ of rapid decrease in $\bf R^{3,1}$ ) is mapped into
linear operators $\lbrace {\cal L}(f) \rbrace \subset  op({\cal H})$ defined on a domain $D$
of vectors dense in ${\cal H}$, independent of $f$ and such that ${\cal L}(f) \ D \subset D$.

\m \n$ Second\ axiom$ : There is a unitary representation $U(a,\Lambda)$ of the Poincar\'e
group such that  $$U(a,\Lambda){\cal L}(f)U^{-1}(a,\Lambda) = {\cal L}(f_{(a,\Lambda)}) ,$$ and
keeping the domain $D$ invariant ($covariance$).

Consider the translation of vector $a$, the spectral decomposition of $U(a,0)$ on ${\cal H}$
is$$ U(a,0) =  \int e^{i(p,a)}dE(p)$$ where $E(p)$ denotes the unic projection-valued spectral 
measure determined by $U(a,\Lambda)$, and $(p,a)$ is the scalar product relative to the Minkovski
metric.The energy-momentum operator is then defined by $P =
\int p.dE(p)$.     \m \n $Third\ axiom$ : $\lbrace p = 0 \rbrace$ is an eigenvalue of $P$ with
multiplicity one and the corresponding eigenspace $\Psi_o \in { \cal D}$ is invariant under all
symmetry transformations, and unique up to a constant phase factor ($uniqueness \ of \ the \
vacuum$). The rest of the spectrum of $P$ lie in or on the plus light-cone ($spectral \ condition$).
\m \n$ Fourth\ axiom$ : If the supports of $f_1$ and $f_2 \in \cal S$ are space-like separated then 
($locality$)$$\lbrack {\cal L}(f_1) , {\cal L}(f_2) \rbrack = 0 .$$ \m \n$ Fifth\ axiom$ : The set
$D_o$  obtained from $\Psi _o$ by applying polynomials in the smeared fields, is a dense set in
${\cal H} (cyclicity
 \ of \ the \ vacuum)$.

Finally to make from the axiomatic field theory a $particule \ theory$, it is necessary to have
in mind the following

\m \n  $ Tacit\ axiom$ : The one-particle asymptotic states can be created from
$\Psi _o$ by some  polynomial  in the smeared fields,
and, in the absence of bound states, the spaces of asymptotic states ($lim_{t\rightarrow
+\infty\ and\ -\infty}$)  coincide with ${\cal H}$ ($asymptotic \ completeness$).

A characteristic feature of the axiomatic field theory is that the quantum field operator does
not exist as an operator defined at each point of spacetime, only smeared fields make sense
mathematically . From this rigorous formalism many efforts have yielded  a number of  by-products
and very general insights into the structure of the theory which concretize, for instance, in
the algebraic approach to QFT,  Ref.[2].  But, except free fields, no interesting models are known.
Then a question  was asked, is it possible to extend this scheme to include the gravitational
quantum field?

\m
\n {\bf 2. The axiomatic approach to quantum gravity} [3]. 

To stay in the spirit of general relativity, it is necessary that the gravitational field
dictates the geometry of spacetime.This implies that the gravitational field is an operator
valued covariant symetric tensor distribution with a canonically associated light-cone field
in order to define the locality and insure the uniqueness of the vacuum. So let us recall
some definitions.

 Let $V$ be a  four-dimensional connected paracompact manifold, a
$cone field \ \cal C$ on $V$ is a mapping which associates to every $x\in V$ a cone ${\cal C}_x
\subset T_x V$ (where cone means the set of isotropic vectors in $x$ for a Lorentzian metric
$g$ said $compatible$ with $\cal C$). A $C^{\infty}$ path is said to be
$physically \ spacelike$ if there is no timelike or isotropic (with respect to $g$)
$C^{\infty}$ path connecting any couple of points of the path. Two points of
$V$ are said to be $ physically \ spacelike \ separated$ if there exists a  physically
spacelike  $C^{\infty}$ path joining the two points, but  two points of $V$ are said
to be $spacelike \ separated$ if they cannot be joined by a timelike or isotropic  $C^{\infty}$
path. Then $g$ is said to be  $strongly \ geodesically \ complete$ if any two timelike or
physically spacelike separated points are joined by a complete geodesic. Denote by 
{\cal S}$^2{\cal D}\left( V \right)$, the set of $ C^\infty $ twice-contravariant
symmetric tensor fields times twisted four-forms on $V$ (following the terminology of Ref.[4])
with compact support and with the $\cal D$-space topology. If $F$ is a topological vector
space over  ${\bf C}$, let {\cal S}$^2{\cal D}'\left( V,F)\right)$ be the set of the linear
continuous applications over {\cal S}$^2{\cal D}\left( V \right)$ with values in $F$. Finally
denote by  ${\cal D}\left( V \right)$ the set of $ C^\infty $ twisted four-forms on $V$ with
compact support. Then  ${\cal G}\in $ {\cal S}$^2{\cal D}'\left( V,F)\right)$ is said an
$F-valued \ distribution \ metric$ if the map ${\cal C}:x\rightarrow{\cal C}_x$ is a cone
field and $\cal C$ will be called the $ cone \ field \ 	associated \ with \ \cal G$.
Moreover the "classical" metric $g$ and the quantum one's $\cal G$ are correlated through  
$${\cal C}\equiv\left\{X\in T V |{\cal G}(X\otimes X\otimes f) = 0,\
\forall f \in{\cal D}(V)\right\}=\left\{X\in T{ V} |g(X,\ X)=0\right\}.\leqno(2.1)$$ Now
a definition for the $gravitational \ quantum \ field$ in an axiomatic form can be given.

 The $axiom \ (0)$ is kept preserved. The $axiom \ (1)$ is also preserved but now the
gra-vitational field is a strongly geodesically complete distribution metric  ${\cal G}
\in{\cal S}^2{\cal D}'\left( V,op({\cal H})\right)$.

 The $axiom \ (2)$ is adapted as follows

\n$ Second\ axiom$ : For every $x \in V$, there is a unitary representation $U_x(a,\Lambda)$ of
the Poincar\'e group such that 
 $$U_x(a,\Lambda) \ {\cal G }{((X\otimes
Y + Y \otimes X) \otimes f)} \ U_x^{-1}(a,\Lambda) = {\cal G }((X\otimes
Y + Y \otimes X) \otimes f_{(a,\Lambda)}),$$
for all $f \in{\cal D}(V) , X,Y \in{T_xV},$ and keeping the domain $D_x = {\cal H}_x\cap D $
 invariant \break ($(local) covariance$).

Now the spectral decomposition of $U_x(a,0)$ on ${\cal H}_x$
is given by$$ U_x(a,0) =  \int e^{i(p,a)_{g_x}}dE_x(p)$$ the integral being taken over $T_x V$.The 
 energy-momentum operator is then defined by $$P_x = \int p.dE_x (p).$$
\indent  Concerning the $axiom \ (3)$ there is just one comment: the vacuum state $\Psi _o$
is invariant under $ U_x(a,\Lambda)$ and $\Psi _o \in D_x$, but is
independent of $x$. Moreover the total spectrum of the energy-momentum operator is equal to
${\cal C}_x$.

The $axiom \ (4)$ is conserved with $f_1$ and $f_2 \in $ {\cal S}$^2{\cal D} \left( V
\right)$.

The $axiom \ (5)$ is unaltered.

The consequences of this formulation, where the gravitational  quantum  field generates its
own background geometry, are studied in Ref.[3]. The first result which stems from the axiom (1)
is the following: $ A \ necessary \ and \ sufficient \ condition \ for \ {\cal G} \ to \ be \
a \ distribution $ 

\n $metric \ is \ that$  $${\cal G}={\cal F}g \ \hbox{o\`u} \ {\cal F}\in{\cal
D}'\left( V,\ op({\cal H})\right)\leqno(2.3)$$ i.e. the gravitational  quantum  field $\cal
G$, being a distribution metric, must be the product of a usual (scalar) operator-valued
distribution  ${\cal F}$ times a classical metric $g$, there is only one quantum degree of
freedom: the spin 0 field ${\cal F}$ !

A second result comes from the covariance axiom (2). It has been shown in Ref.[3] that: $the \
covariance \ principle \ is \ valid \ if \ and \ only \ if \ g \ is \ conformal \ to \ a \
flat \ metric.$ 

But the covariance axiom must be modified, indeed the Poincar\'e group appears in relativity
as the isometry group of the Minkovski space, and, in the above framework, we have to take
into account for the presence of the "classical" metric. This can be done by prescribing the
covariance under the isomorphism group of the "classic" geometrical structure, as advocated
by the authors of  Ref.[5], but with the restriction that it exists at least a one-dimensional
translation subgroup to be able to define the vacuum state. So amended, the above axiomatic
approach leads to a description of the quantum gravity as the semiclassical theory of a scalar
quantum field in a background Lorentzian manifold, i.e. as a semiclassical theory, widely
studied, see for instance  Ref.[6], but it is generally not believed that this framework can provide
an exact description of nature.

 In  Sect.3, a geometrical structure is exhibited
which goes in the same way as the above "amended" axiomatic, then an interpretation of the
very early quantum cosmology is proposed in Sect.4. \vfill \eject

\m
\n {\bf 3. The geometry of isotropic hypersurfaces ($ISHYPS$) foliated Lorentzian manifolds and the
Einstein's field equations} [7].

Let us consider a Lorentzian manifold ($V,g$) endowed with an isotropic vector field $\xi$, i.e.
$g(\xi,\xi) = 0$, and a one-dimensional family  $\lbrace \Sigma \rbrace$ of $ishyps$ such that the
"degenerate" metric $\beta$ induced by $g$ on each $\Sigma$ has its kernel generated by the line
field  $[\xi]=\left\{ \lambda\xi , \lambda\in{\bf R}-\{0\}|\beta(\xi,.)=0\right\}.$

Then, at each point $x$ of any $\Sigma$, the cone of isotropic directions has a first order contact
with $\Sigma$ but the tangent space $T_x\Sigma$ does not contain any time-like vector, the future
light cone of $x$ being entirely on one side of $\Sigma$, and the past light cone entirely on the
other side.

Let $Gl(V)$ be the principal fiber bundle of linear frames on $V$. Then the presence of the metric
tensor $g$ leads to the reduction of  $Gl(V)$ to the bundle of orthonormal frames $O(V)$, moreover
we have to keep fixed the line field, which introduces an other subbundle $G(V)$,the structure
group $G$ of which is a semi-direct product     ${\bf R}^2 \oslash {(SO(1,1) \otimes
SO(2))}$. Therefore $G(V)$ is the $adapted$ frame bundle to the triplet ($V,g,[\xi]$). However a
notion of $teleparallelism$ traces back to Weizenbock (1923) and was introduced into physics by
Einstein, Ref.[8], in which is brought into play a connection into an other subbundle denoted by
$G_T(V)$, where $G_T = SO(2) \subset G$. Now, if one looks for subbundles
of $G(V)$ which contain $G_T(V)$, two possibilities appear. Firstly a subbundle with structure group
$G_I = {\bf R}^2 \oslash  {SO(2)}$ with the property that the reduced canonical connection parallel
transports the vector field $\xi$, then generalized $pp$-waves are an example of spacetimes admitting
such a reduction. Secondly a subbundle $G_R(V)$ the property of which being that the reduced
canonical connection is a connection on any $ishyps$ $ \Sigma $. It has been denamed $the \ 
radiation-connection$ and $G_R = SO(1,1) \otimes SO(2)$. Underlying spacetime manifolds, 
compatible with such a reduction, are isometric to the direct product  $V_{2,0} \times V_{1,1}$ of
two bidimensional spaces, one of them $V_{2,0}$ being space-like and Riemannian, the other one $
V_{1,1}$ being time-like and Lorentzian, see Ref.[9]. Then any $ishyps \ \Sigma$ can be viewed as
$\Sigma =
 V_{2,0} \times \gamma$ where $\gamma$ is an integral curve of the vector field $\xi$, light-like in
$ V_{1,1}$ as well as in $V$.

Now, let us examine the impact of the radiation connection over the Einstein's field equations
$\cal E = \cal T$, more precisely over the right hand side of the equations, i.e. the $physical$
tensor $\cal T$, since the Einstein's tensor $\cal E$ is strongly constrained by the above described
product structure of the spacetime.

It has been shown, in Ref.[9], that the 2-covariant tensor $\cal T$ is characterized by only two
suitably differentiable functions over $V$. This two functions, denoted by $\chi$ and
$\lambda$, are such that $$\normalbaselineskip=18pt\left\{\matrix{(a)&Tr_g \ {\cal
E}&=&\chi\hfill\cr (b)&{\cal E}(\xi)&=&\lambda \  g(\xi)\cr}\right.\leqno(3.1)$$The first relation
shows that $\chi=-S,\ S$ denoting the scalar curvature of $(V,g)$, and, from the second, one deduces
that the light-like vector field is solution of the equation $\xi={1\over\lambda} \ {\cal E}(\xi), \
{\cal E}$ denoting the mixed tensor here.

Then, by choosing an isotropic 1-form $\theta$ (i.e.   $g^{-1}\left(\theta,
\theta\right)=0$ ) such that $\xi\rfloor\theta=1$, $g^{-1}(\theta)$ is an
isotropic vector field transversal in each point of $V$ to the $ishyps$ which contains the point,
and the $physical$ 2-covariant tensor $\cal T$ can be written as $${\cal T}=\left({\chi\over
2}-\lambda\right)g-\left({\chi\over2}-2\lambda\right)\left(g(\xi)\otimes\theta+\theta\otimes
g(\xi)\right). \leqno(3.2)$$ \indent  From the above expression, a standard familly of spacetimes
admitting the $radiation$ \

\n $-connection$ clearly apppears, they are the $ Einstein's \ spacetimes$
obtained by taking  $$\chi=4\lambda=-S= constante,$$  so ${\cal T}=\lambda g \ \left(=-{1\over 4}S
g\right)$. From this remark, it is suggested that $\chi$ and $\lambda$ can be interpreted as
$cosmological \ functions$. 

Then, in Ref.[7], it was proposed to associate the above described geometrical framework to the
inflationary period of the Universe and to speak of $primordial \ gravitational \ waves$. Moreover,
it should be noticed that a fluid of massless particules moving along $\xi$ cannot be described
by the expression (2.1) of ${\cal T}$. Hence we are faced with a geometrical scheme from which
gravitons are manifestly excluded, a situation already met in the axiomatic approach to quantum 
gravity, whence the idea to bring together both frameworks.

\m
\n {\bf 4. The quantum scalar field confronted with the radiation structure} [10].

\n
The dynamics for the $\cal F$ field does not proceed from the axiomatic described in Sect.{\bf 3},
so it has to be postulated. By taking into account for the preponderant part of $ishyps$ into our
investigation, we are led to propose an $isotropic$ dynamics for $\cal F$ (something like the
Dirac's $light \ front \ form$). For every $ishyps \ \Sigma$, one considers the algebra ${\cal A}_{\Sigma}$
 generated by the physical observables  measured in $\Sigma$. The dynamics is then defined by
$\star$-isomorphisms   ${\cal A}_{\Sigma}\to{\cal A}_{\Sigma'}$ deduced from field equations. Let
us note that the causality condition which tells us that everything we can quantize in the metric
tensor is the scalar factor $\cal F$, takes a particular aspect in our approach. There is a causal
dependence between two connected sets  ${\cal O}$ and ${\cal O'}$ belonging to the same isotropic
geodesic. If  ${\cal O}$ is in the future of ${\cal O'}$, then ${\cal A}({\cal O})\subset{\cal
A}({\cal O'})$. On the other hand if  ${\cal O}$ and ${\cal O'}$ belong to two different isotropic
geodesics respectively, they are spacelike separated and the $locality \  property$ is recovered [11]
$$\left[{\cal A}({\cal O}),\ {\cal A}({\cal O'})\right]=0. \leqno(4.1)$$
At this stage we must adhere to the principle of simplicity by using natural objects of the
radiation structure only, then, tentatively, we may write the field equations as follows

\n $Dynamical \ "axiom"$ : The classical background spacetime  $(V,g,\xi)$ is isometric to $V_{2,0}
\times V_{1,1}$
 which satisfies the Einstein's field equations with the following $physical$ tensor  $${\cal
T}=\left({\chi\over
2}-\lambda\right)g-\left({\chi\over2}-2\lambda\right)\left(g(\xi)\otimes\theta+\theta\otimes
g(\xi)\right). \leqno(3.2)$$ 

The quantum factor of the metric is minimaly coupled to the classical background through the
differential system $$\normalbaselineskip=18pt\left\{\matrix{(a)& \widetilde {\xi}({\cal
F})&=&\kappa{\cal F}\cr (b)&\widetilde{\Delta}{\cal F}\hfil&=&0\hfill\cr}\right.\leqno(4.2)$$
\n where $\kappa \in \bf C$.

\m Obviously the differential operators  $\widetilde {\xi}$ and $\widetilde{\Delta}$ being
defined by duality, Rel.(4.2 a-b) mean that
$$\normalbaselineskip=18pt\left\{\matrix{(a)&<\widetilde {\xi}{\cal F},\ f>&=&-<{\cal
F},\ \xi(f)>&=&<{\cal F},\ \kappa f>\cr
(b)&<\widetilde{\Delta}{\cal F},\ f>&=&<{\cal F},\  \Delta_{conf}
 (f)>&=&0.\hfill\cr}\right.\leqno(4.3)$$Hence the behaviour of $\cal F$ along the
vector field $\xi$  is governed by Rel.(4.3 a) with the eigenvalue $\kappa$ to be interpreted. In
Rel.(4.3 b), the Laplacian associated to the conformal class of $g$ is denoted by $\Delta_{conf}$
and is defined as usually by $$\Delta_{conf}=\Delta_g+{1\over 6} \ S\leqno(4.4)$$
where the standard Laplacian associated to $g$ is given by  $$\Delta_g=g^{-1}(\nabla_R ,d) ,
\leqno(4.5)$$  $\nabla_R$ denoting the covariant derivative associated to the radiation connection,
and $d$ the external derivative.

However consequences of the $G_R$-reduction have not yet been fully exploited. The isotropic 1-form
$\theta$ can be used to construct a quasi-inverse $\beta_{\theta}$ to the first fundamental form 
 $\beta$ induced by  $g$ over any $\Sigma$. This quasi-inverse is defined by the property that the
contraction of $\beta$ with $\beta\beta_{\theta}$ gives again $\beta_{\theta}$, which leads to
$$\beta\beta_{\theta}={\bf1}-\theta\otimes\xi.\leqno(4.6)$$
Then, over any $\Sigma$, a Laplacian denoted by $\Delta_{\beta}$ can be defined
$$\Delta_{\beta}=\beta_{\theta}\left(\nabla_{R} , d \right).
\leqno(4.7)$$
Modulo the choice of  $\theta$, the Laplacian
$\Delta_g$ can be splitted up under the following form
$$\Delta_g=\Delta_{\beta}+2({\nabla}_{R})_{g^{-1}(\theta)}
\otimes\xi.\leqno(4.8)$$

By taking into account for this decomposition of $\Delta_g$ the differential system (4.3) is
replaced by a parabolic type differential equation $$B^{\star}_g f:\
=\left(\kappa ({\nabla}_{R})_{g^{-1}(\theta)}-{1\over
2}\left(\Delta_{\beta}+{1\over 6}S\right)\right)\, f\,=0.\leqno(4.9)$$

But the operator $B^{\star}_g$ can be considered as the adjoint of a differential operator
 $B_g$ defined over the  distributions of $V$ as follows
$$<B_g{\cal F},\ f>=<{\cal F},\ B^{\star}_g f>\quad\forall{\cal F}\in{\cal
D'}( V) \ \hbox{et}\ f\in C^{\infty}_{\circ}( V).\leqno(4.10)$$

Then $B_g$ can be inserted into the system (4.2), and one gets a parabolic type dynamic for the
quantum field operator ${\cal F}$ given by the homogeneous equation $$B_g{\cal
F}=0\leqno(4.11)$$ which controls the evolution of ${\cal F}$ along the isotropic vector field
$g^{-1}(\theta)$ which points out of   $\Sigma$. \vfill \eject
$Remarks$

\n  (1) - Existence and uniqueness theorems are known about quantum field theory of null
hyperplanes, but in the Minkovski space (see for instance  Ref.[12]).

\n  (2) - In Ref.[13], it is shown that the wave equation always admits one and only one elementary
solution which furnishes the distribution solution of a Cauchy problem by a composition with the
initial data but on a globally hyperbolic manifold.

\n  (3) - Let us return to the proposed dynamics in its initial form. Rel.(4.2 b) is a particular
case of the scalar wave equation studied in Ref.[14], the theorem (4.5.1) of which asserts the
existence of fundamental solutions. Then, according to Ref.[13], the solutions of the homogeneous
equation (4.2 b) can be obtained by convoluting the propagator with a scalar distribution.
Finally one has to select among these solutions, the ones (if they exist) which satisfy the
eigenvalue equation (4.2 a).
\m \n {\bf By way of conclusion: a speculation about the quantum origin of the Universe.}
\bigskip  Let us try to insert into a cosmological scenario the above described scheme.  At large
scale, the Universe, such as one  experiences it, is classic and well interpreted by general relativity.
Otherwise the scientific community is ready to have been persuaded of the quantum origin of the
Universe. In such a case we have to acknowledge for the existence of a spell through which quantum
described phenomenons, such as matter creation, are attended by the classical spacetime creation,
the spacetime in the heart of which the matter settles down to result in the Universe we belong
to. This spell can agree with the inflation time widely advocated to-day. In the course of this
brief (less than $10^{-36} $ sec.)  interlude, it is predicted that every spacelike area, occupied by
the primordial matter, increased by a $10^{100}$ scale factor before to reach its classical
expansion speed, during which only small scale quantum effects take place. 

Therefore the quantum axiomatic, grounded on the uniqueness of the cone field in classical
and quantum formalism as detailed in Sect.2, seems well adapted to treat at once phenomenons
hinging on both descriptions. To explain how the quantum field $\cal F$ can be matter engendring,
we shall base arguments upon dimensional analysis of the described dynamics in Sect.4. If in
Rel.(4.9), the dimension  $[L^{-2}]$ is attributed to the spacial Laplacian $\Delta_{\beta}$, and if
one decides to measure the quantum evolution with a classical time unit by imposing the dimension $[T^{-1}]$   
to $ ({\nabla}_{R})_{g^{-1}(\theta)}$ one gets $[\kappa]=[L^{-2}T]$, the dimension of a mass
quotiented by an action. So it is natural to suppose that $\kappa$ is proportional to $m/\hbar$,
and to set $\kappa=i{m\over\hbar}$. Hence the quantum evolution is piloted by the Schr\oe dinger
equation (4.11) and follows an isotropic direction of the budding classical spacetime.

But it should be noticed that the original tensor $\cal T$ is modified by the appearance  of matter
created by the quantumfield. However it is allowable to suppose that the to-day "observed" little
cosmological constant is the relic of the primordial tensor $\cal T$ from which the initial
structure ("past horizon") of our (at large classical) spacetime has been shaped. 
\vfill \eject

\n {\bf References}
\bigskip

\parindent=1cm

\item{\hbox to\parindent{\enskip [1]}\hfill}R.F. STREATER and A.S. WIGHTMAN,  {\it PCT, Spin, 
Statistics, and all that}, Benjamin Inc. N.Y., (1964).
\medskip
\item{\hbox to\parindent{\enskip [2]\hfill}}R. HAAG, D. KASTLER,J.Math.Phys.{\bf 5} , 848-861 
(1964). \item{\hbox to\parindent{\enskip [3]\hfill}}M. FLATO, J. SIMON, Physica Scripta  {\bf
3}, 53-59 (1971). \item{\hbox to\parindent{\enskip \hfill}}M. FLATO, J. SIMON, Phys. Rev.
D {\bf 5}, 332-341 (1972).
\item{\hbox to\parindent{\enskip \hfill}}M. FLATO, J. SIMON, D.
STERNHEIMER, Phys. Rev. Letters {\bf 28}, 385-387 (1972).
\item{\hbox to\parindent{\enskip \hfill}}M. FLATO, J. SIMON, D.
STERNHEIMER, Helv. Phys. Acta {\bf 47}, 114-118 (1974).
\item{\hbox to\parindent{\enskip [4]\hfill}}G.de RHAM,{\it  Vari\'et\'es
diff\'erentiables}, Hermann Ed. Paris (1955). \item{\hbox to\parindent{\enskip \hfill}}L. SCHWARTZ,
{\it Th\'eorie des distributions},  Hermann Ed. Paris (1966).

\item{\hbox to\parindent{\enskip [5]\hfill}}B. de WITT, Phys. Rep. {\bf 19 C}, 6, 297-357 (1975)
\item{\hbox to\parindent{\enskip \hfill}}J. DIMOCK, Commun. Math. Phys. {\bf 77}, 219-228 (1980).
\item{\hbox to\parindent{\enskip [6]\hfill}}A. ASHTEKAR, A. MAGNON, Proc.Roy.Soc. A {\bf 346},
375-394 (1975). \item{\hbox to\parindent{\enskip \hfill}} P. HAJICEK, in {\it  Differential
geometrical methods in mathematical physics II} (eds K.Bleuler,H.Petry, A.Reetz), 535-566, Springer
(1978). \item{\hbox to\parindent{\enskip \hfill}}C.ISHAM, 459-512, -$idem$- .
\item{\hbox to\parindent{\enskip [7]\hfill}}G.BURDET,  M.PERRIN,   Monographie Erasmus
{\it  "Mathematics and Fundamental Applications "}  {\bf 50} (1992).
\item{\hbox to\parindent{\enskip \hfill}}G.BURDET,T. PAPAKOSTAS, M.PERRIN, 
	 General Relativity and Gravitation {\bf 26}, 225-232  (1994) and 
Int. Journal Modern Physics D. {\bf 3}, 163-16. (1994).
\item{\hbox to\parindent{\enskip [8]\hfill}}R. WEINTZENBOCK, {\it Invariantentheorie}, Moordhoff
Groningen, (1923). \item{\hbox to\parindent{\enskip \hfill}}A. EINSTEIN, {\it Sitzungsber. Akad.
Wiss.} (Berlin),  Phys.-Math. Kl, {\bf 217}, 224-227 (1928).
\item{\hbox to\parindent{\enskip [9]\hfill}}G. BURDET, M.PERRIN, Lett. Math. Phys. {\bf 25}, 39-4.
(1992). \item{\hbox to\parindent{\enskip [10]\hfill}}G. BURDET, M.PERRIN, Lett. Math. Phys. {\bf
30}, 317-3.. (1994).
\item{\hbox to\parindent{\enskip [11]\hfill}}R. HAAG, {\it Local Quantum
Physics}, Texts and Monographs in Physics, Springer Verlag, (1992).
\item{\hbox to\parindent{\enskip [12]\hfill}}H. LEUTWYLER, J.R. KLAUDER, L. STREIT, Nuov.Cim. {\bf
LXVI A}, 536-554 (1970).
 \item{\hbox to\parindent{\enskip \hfill}}R. A. NEVILLE, F. ROHRLICH, Nuov.Cim. {\bf 1 A},
625-643 (1971).
 \item{\hbox to\parindent{\enskip [13]\hfill}} A. LICHNEROWICZ, {\it Propagateurs et Commutateurs en
Relativit\'e G\'en\'erale}, Publications Math\'ematiques I.H.E.S. n$^{\circ}$10 (1960),
\item{\hbox to\parindent{\enskip \hfill}} Y. CHOQUET-BRUHAT, {\it Hyperbolic
Partial Differential Equations on a Manifold}, Battelles Rencontres 1967, Ed.
by  C.M. de WITT et J.A. WHEELER, p.~84-106, Benjamin Inc. Voir aussi Y. BRUHAT,
Annali di Matematica Pura ed Applicata, {\bf 64}, 191-228 (1964).
 \item{\hbox to\parindent{\enskip [14]\hfill}}F.G. FRIEDLANDER, {\it The wave
equation on a curved space-time}, Cambridge University Press (1975).

 \end